# Implementation of field two-way quantum synchronization of distant clocks across a 7 km deployed fiber link

Runai Quan[1,2], Huibo Hong[1,2], Wenxiang Xue[1,2], Honglei Quan[1,2], Wenyu Zhao[1,2], Xiao Xiang[1,2], Yuting Liu[1], Mingtao Cao[1], Tao Liu[1,2], Shougang Zhang[1,2,+], Ruifang Dong[1,2,*]

[1] Key Laboratory of Time and Frequency Primary Standards, National Time Service Center, Chinese Academy of Sciences, Xi'an, 710600, China

[2] School of Astronomy and Space Science, University of Chinese Academy of Sciences, Beijing, 100049, China

*E-mail: dongruifang@ntsc.ac.cn, szhang@ntsc.ac.cn

**Abstract:**
**The two-way quantum clock synchronization has been shown not only providing femtosecond-level synchronization capability but also security against symmetric delay attacks, thus becoming a prospective method to compare and synchronize distant clocks with both enhanced precision and security. In this letter, a field test of two-way quantum synchronization between a H-maser and a Rb clock linked by a 7 km-long deployed fiber was implemented. Limited by the frequency stability of the Rb clock, the achieved time stability at 30 s was measured as 32 ps. By applying a fiber-optic microwave frequency transfer technology, the stability was improved by more than one-magnitude to 1.9 ps, even though the number of acquired photon pairs was only 1440 in 30 s due to the low sampling rate of the utilized coincidence measurement system. Such implementation demonstrates the high practicability of two-way quantum clock synchronization method for promoting the field applications.**

## Introduction

Precise synchronization between distant clocks is of great significance due to its essential role in almost every type of precision measurement. The two-way time transfer is a popular way for precise comparison and synchronization between distant time scales. Among different two-way time transfer (TWTT) methods, the widely utilized satellite-based TWTT has reached a time stability of 200ps [1,2]. With a much higher frequency and bandwidth than radio radiations, the optical TWTT allows realizing time stability of a few picoseconds and accuracy better than 100 ps [3-6]. Benefitting from the low loss, high reliability, and high stability of optical fibers, the fiber-optic TWTT offers an alternative method to further improve the precision [7,8]. With the rapid development of atomic clocks [9,10], new method requires to be developed for potential higher-precision clock synchronization applications.

By virtue of the strong temporal correlation characteristics of time-energy entangled photon pair sources, fiber-based two-way quantum clock synchronization protocol was proposed to compare and synchronize the distant clocks [11]. With the common frequency reference and epoch time standards, the femtosecond-level synchronization stability was demonstrated over a 20-km optical fiber link, together with a much-enhanced accuracy of 2.46 ps [12]. At the same period, the proof-of-principle synchronization experiment between different frequency references was also reported. With two independent Rb clocks as the references, a time stability of 51 ps at 100 s was achieved, which was limited to the intrinsic instability of the Rb clock [13]. To tackle the practical synchronization of distant clocks and explore its possible advancement, it is imperative to extend the two-way quantum synchronization method to field tests.

In this paper, we carried out a field test of the two-way quantum clock synchronization between a H-maser located at the campus of National Time Service Center (NTSC) and a Rb clock at the Lishan Observatory (LSO) of NTSC. Through the deployed telecommunication fiber with a distance of 7 km, the two sites were physically connected. Similar with the report in Ref. [13], the reached time stability of 32 ps at 30 s was restrained by the Rb clock. By introducing the fiber-based microwave frequency transfer into the experiment, the stability was improved by more than one order of magnitude to 1.9 ps with only 1440 coincidence counts in 30 second, which was due to the low sampling rates of the utilized event timers (ET, ~6 kHz) for coincidence measurement. Further improvement of the time stability is highly expected by utilizing new ETs with MHz-level sampling rate [14]. This experimental demonstration lays a solid foundation for the application of two-way quantum clock synchronization in practical systems for potential sub-picosecond precision. Benefitting from it, time-bin-encoded quantum key distribution can also be implemented without additional dedicated synchronization procedure [15,16].

**Experimental Setup**

The schematic diagram of the field two-way quantum clock synchronization setup is shown in Fig. 1. The H-maser sited at the NTSC lab and the Rb clock (PRS10, SRS. Inc) at the LSO lab were connected by a 7 km-long deployed fiber link. At each site, there is a time-energy entangled photon-pair source, a pair of single photon detectors (SPDs) and an ET referenced to the local clock. The photon pair sources were generated by using a 780 nm DBR laser (Photodigm Inc) to pump a 10 mm-long periodically poled Lithium Niobate (PPLN, type II) waveguide with the poling period of ~8.4 μm. Via the spontaneous parametric down conversion (SPDC), degenerate photon pairs at 1560 nm (denoted as signal and idler photons) were obtained with orthogonal polarizations [17]. The four SPDs were superconductive nanowire single photon detectors with efficiency of 65% (SNSPD, Photec Ltd.) [18,19], denoted as D1-D4. Two commercial ETs (A033-ET, Eventech Ltd), each having two input ports and a sampling rate of 6 kHz for one port, were utilized as the time tagging devices. At NTSC, the signal photon ($s_1$) was kept for local detection by D1, and the idler photon ($i_1$) was transmitted forward through the 7 km fiber to LSO to be detected by D2. Similarly, $s_2$ was maintained locally at LSO and detected by D4, while $i_2$ was transmitted backward to NTSC and detected by D3. The optical circulators (OC1 and OC2) were used to ensure the bidirectional transmission through the same fiber.

The photons detected by D1, D2, D3, D4 were subsequently tagged by the local ETs as $\{t_1^j\}, \{t_2^j\}, \{t_3^j\}, \{t_4^j\}$, where the superscript $j$ denotes the $j$-th tagged photon. The time tags $\{t_2^j\}, \{t_3^j\}$ at LSO were sent to NTSC via classical communication channel for acquiring the time differences $t_2 - t_1$ and $t_3 - t_4$ by nonlocal coincidence identification algorithm [20]. According to the bidirectional symmetry, the time offset $t_0$ between the two clocks can be extracted by $((t_2 - t_1) - (t_3 - t_4))/2$ [12].

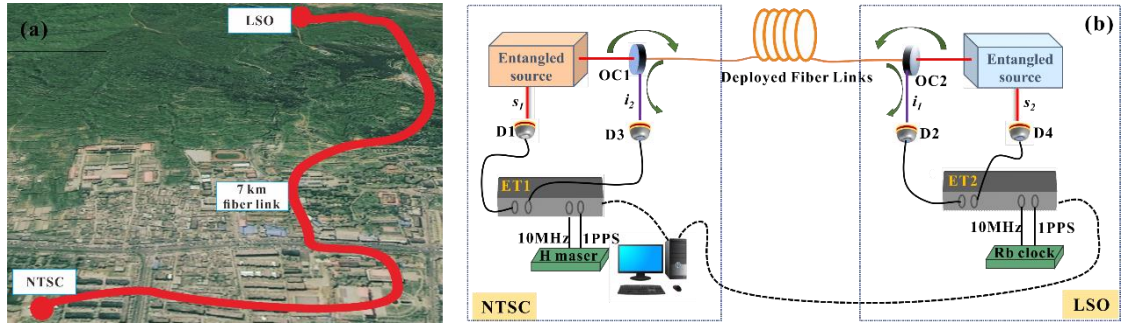

Fig. 1 (a) Aerial and (b) schematic view of the experimental setup of the field two-way quantum clock synchronization between the H-maser at NTSC and the Rb clock at LSO.

**Experimental Results**

1. Evaluation of synchronization stability dependent on the reference clocks

The influence of the reference clocks on the synchronization stability was firstly analyzed by setting the two-way quantum clock synchronization setup locally in the NTSC lab and using a 3-m fiber to link the two sites. Since each reference clock offered a 10 MHz and a 1 PPS signal to the ET for frequency reference and epoch time standards, four cases of settings were tested: case 1) both ETs were referenced to the same 10 MHz and 1 PPS from the H-maser, case 2) both ETs were referenced to the same 10 MHz from the H-maser, but the 1 PPS signals for them were individually from the H-maser and the Rb clock, case 3) both ETs were referenced to the same 1 PPS from H-maser, but the 10 MHz signals were individually from H-maser and Rb clock, and case 4) the two ETs were respectively referenced to H-maser and Rb clock. The corresponding time offset results versus the elapsed time of about 190 s are presented in Fig. 2. It can be seen that, with the same 10 MHz reference frequency, the measured time offsets remain independent on the elapsed time (in wine stars and blue up-triangles) even when the referenced PPS signals are different; while for different reference frequencies, the measured time offsets experience significant drift with respect to the elapsed time (in black squares and magenta down-triangles).

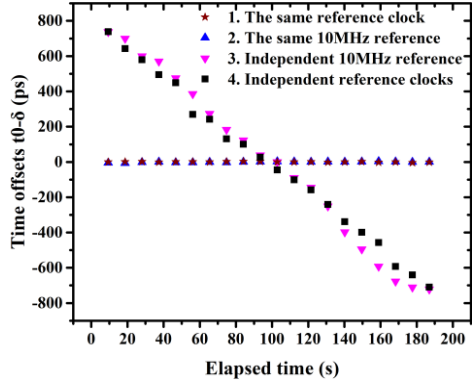

Fig. 2 The time offset results acquired under the cases of 1) the same reference clock (in wine stars), 2) the same 10MHz reference (in blue up-triangles), 3) different 10MHz frequency references (in magenta down-triangles), and 4) different reference clocks (in black squares).

For the case with independent reference clocks, a long-term time offset measurement was implemented and Fig. 3 (a) depicts the 2-hour results in black squares. By applying a quadratic polynomial fitting to the result [21], the relative frequency drift between the H-maser and the Rb clock is given as $\sim 1\times 10^{-11}$, which coincides well with the nominal relative frequency accuracy ($<5\times 10^{-11}$) of the utilized Rb clock. The residuals of the measured time offsets after the quadratic polynomial fitting are shown in Fig. 3 (b), which shows a fluctuation of about 162.6 ps in standard deviation. The attainable synchronization stability in terms of Allan deviation (ADEV) is analyzed and given by blue diamonds in Fig. 3 (c), which reaches $2.2\times 10^{-12}$ at an averaging time of 30 s. For comparison, the relative frequency stability of the Rb clock with respect to the H maser was measured by a phase noise analyzer (Symmetricom 5125A) and plotted in Fig. 3 (c) by black squares. The consistency between the two curves within the averaging times of 30-400 s implies that the synchronization performance of our protocol has arrived at the fundamental limit of the clock itself. The inflections around 1000 s in both ADEV curves show consistence with the fluctuation period of the NTSC lab's temperature (shown in Fig. 3 (d)). The upward trend of the blue-diamond curve beyond 1000 s is from the relative frequency drift of the Rb clock. As such frequency drift was removed when evaluating the synchronization stability, the ADEV curve given by black squares remains the downward trend.

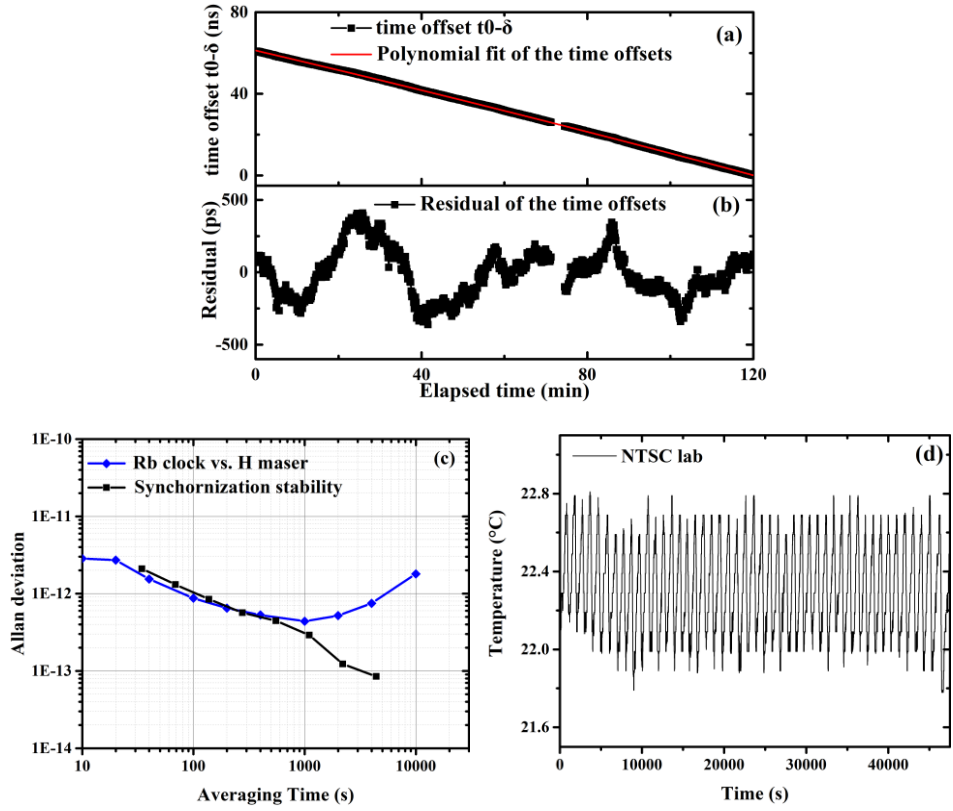

Fig. 3 Evaluated results of the synchronization performance under independent reference clocks. (a) The measured time offsets versus the elapsed time in two hours (black squares) and the corresponding polynomial fit (red line). (b) The residuals of the time offset results as a function of the elapsed time. (c) The attainable ADEV synchronization stability (black squares), and the relative frequency stability of Rb clock to H maser (blue diamonds). (d) Part of the recorded temperature fluctuation of the NTSC lab during the measurement.

2. Evaluation of the ETs' time tagging performance

To evaluate the time tagging performance of the two utilized ETs, the H maser was used as the common frequency reference for both of them. The signal generator (AFG31052, Tektronix. Inc), which was also referenced to the H maser, was used to generate two 1 PPS signals with a preset relative delay ($\tau$) between them. By individually sending them to the two input ports (denoted by A, B) of ET1 and ET2, the measured time offsets as a function of the preset $\tau$ for both ETs are shown in Fig. 4 (a) in black squares and red circles respectively. In the whole range of measured time offsets, the consistency of the two curves and their preeminent linearity with the preset time value $\tau$ can be clearly seen. To zoom in the plots, there is a fixed difference of 19.7 ± 4.8 ps between the measured time offsets of ET1 and that of ET2, which can be explained as the slight inhomogeneity between the input ports of individual ET.

Subsequently, the homogeneity between the two ETs is also evaluated by sending the two 1 PPS signals with a fixed delay of 100.147 ns into either two of the input ports of them. Therefore, six cases are considered: (1) input A of ET1 as the start, input A of ET2 as the stop (1A2A), (2) input A of ET1 as the start, input B of ET2 as the stop (1A2B), (3) input B of ET1 as the start, input A of ET2 as the stop (1B2A), (4) input B of ET1 as the start, input B of ET2 as the stop (1B2B), (5) input A of ET1 as the start, input B of ET1 as the stop (1A1B), and (6) input A of ET2 as the start, input B of ET2 as the stop (2A2B). The correspondingly measured time offsets are shown in Fig. 4 (b) by black squares, red circles, blue up-triangles, purple down-triangles, olive diamonds, and navy hexagons respectively, which provide the statistical values of 101.034±0.009 ns for 1A2A, 101.097±0.007 ns for 1A2B, 100.984±0.008 ns for 1B2A, 101.045±0.008 ns for 1B2B, 100.196±0.007 ns for 1A1B, and 100.211±0.007 ns for 2A2B. From the above analysis, it can be deduced that a maximal time bias of about 893 ps will be introduced by the ETs. In the absolute time offset evaluation, these bias contributions should be carefully corrected. Meanwhile, each sequence of the time offset results shows a similar standard deviation of 7~9 ps, manifesting the reliability of the two ETs in the time offset identification.

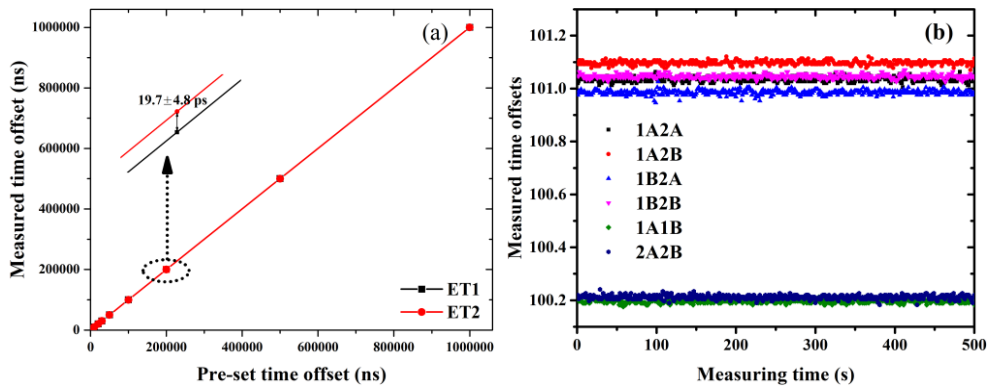

Fig. 4 (a) Measured time offsets of ET1 (black squares) and ET2 (red circles) as a function of the varying time delay between the two input 1PPS signals. (b) Under the fixed time delay of 100.147 ns, measured time offset results with time under different input configurations of 1A2A (in black squares), 1A2B (in red circles), 1B2A (in blue up-triangles), 1B2B (in purple down-triangles), 1A1B (in olive diamonds), and 2A2B (in navy hexagons).

3. Field test result of the quantum F-TWTT setup

Then the two-way quantum clock synchronization between the H-maser in the NTSC lab and the Rb clock in the LSO lab were performed. Based on the experimental system, the coincidence widths of the time differences $t_2 - t_1$ and $t_3 - t_4$, were measured to be around 285 ps and 286 ps respectively, which are consistent with the dispersion broadening of the fiber link. The extracted time offset $t_0$ between the two clocks as a function of the elapsed time is plotted in Fig. 5 (a) by black squares. By applying a quadratic polynomial fitting (red curve), the relative frequency drift between the H-maser and the Rb clock is evaluated as ~ $7.1 \times 10^{-11}$, whose deviation from the previously evaluated result of $1 \times 10^{-11}$ should be from the field fiber link. The residuals of the fit are depicted in Fig. 5 (b) and a breathing fluctuation is shown, which can be attributed to the independent air-conditioning environments at the two sites as their temperature variations feature different periodic behavior.

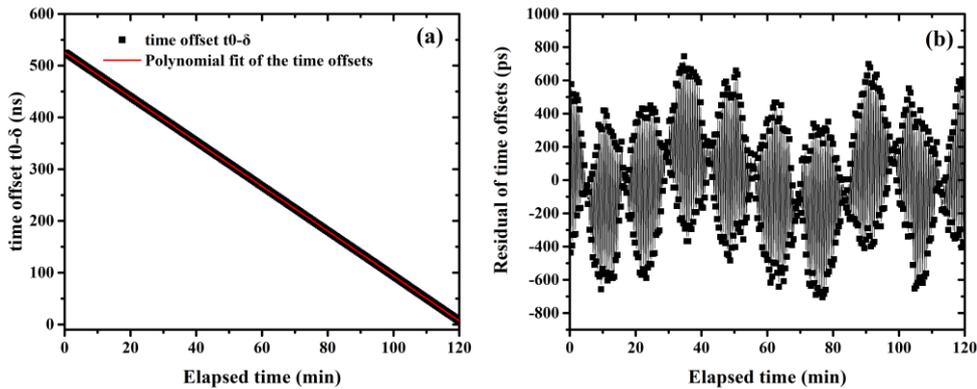

Fig. 5 (a) Measured time offsets between the two remote clocks and (b) the residual of the polynomial fit as a function of the elapsed time.

The synchronization stability between the two distant clocks in time deviation (TDEV) is also evaluated and plotted in Fig. 6 by black squares, in which a value of 38.1 ps at 30 s is obtained. In order to eliminate the dispersion impact in transmission paths, one dispersion compensation fiber (DCF) with a length of 1.245 km was inserted in each local arm and the narrowed coincidence widths about 121 ps were expected due to the nonlocal dispersion cancellation (NDC) [22]. However, the coincidence widths were measured as 177 ps for $t_2 - t_1$ and 180 ps for $t_3 - t_4$ respectively, which are much broader than the simulation and should be attributed to the relative frequency drift between the two clocks. Under NDC configuration, the achieved time stability is also shown in Fig. 6 (red circles) and gives a TDEV of 32 ps at 30 s. We can see that the NDC has trivial influence on the short-term stability. Comparing the long-term time stabilities between the cases with and without NDC in the setup, the almost same results of 19.3 ps at 7680 s further indicate that the NDC has negligible impact on the achieved stability. For clarity, the TDEV result of the two-way quantum clock synchronization setup co-located in the NTSC lab is also given in Fig. 6 by blue down-triangles. It can be seen that, all the TDEV curves first rise with the increment of averaging time and then behave drop-trend. The equivalence of the three TDEV curves before the averaging time reaching 1000s shows that the attainable short-term synchronization stability in field test is also limited by the relative frequency stability of the Rb clock. With the increase of averaging time, the synchronization stability starts to depend more on the ambient fluctuations. In comparison with the case with the two reference clocks at the same lab (blue down-triangles), whose inflexion arrives at 1000 s, the inflexion points with the two clocks located at different labs move to around 2000 s. This movement exactly reflects the independent periodical temperature variation of individual laboratory.

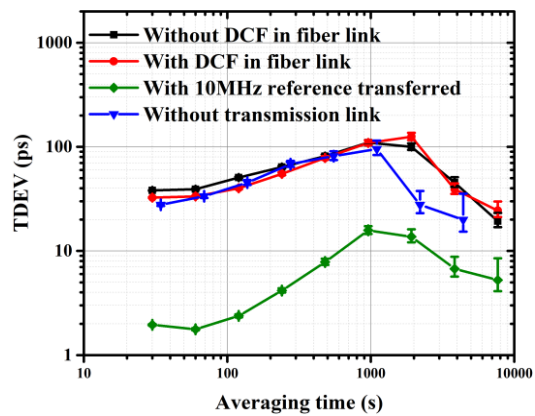

Fig. 6 Synchronization stability between two independent clocks separated by 7 km deployed fibers without DCF in fiber link (black squares), with DCF in fiber link (red

circles), and without 7 km deployed fibers (blue down-triangles); two-way quantum clock synchronization stability with transferring the reference frequency of the H-maser to LSO laboratory (olive diamonds).

4. Synchronization stability improvement with the microwave frequency transfer

As it has been shown that the relative frequency stability of the reference clocks dominates the clock synchronization performance, a distinct improvement can be expected by applying stabilized 10 MHz reference frequency between the two distant sites. In view of that, utilizing a fiber-based microwave frequency transfer technology, the 10 MHz frequency reference from the H maser at NTSC was stably transferred with a stability of $3.7\times10^{-15}$@30 s and used as the ET frequency reference at LSO. Detailed frequency transfer experimental configuration and results are shown in Ref. [23]. Under this accomplishment, the two-way quantum clock synchronization was performed and the resultant time stability is presented in Fig. 6 by olive diamonds. The short-term stability result gives a TDEV of 1.9 ps and a corresponding ADEV of $1.1\times10^{-13}$ at 30 s, which shows a distinguished improvement by more than one order of magnitude. To compare with the above achieved frequency transfer stability, there is a gap of almost two orders of magnitude to be analyzed. According to the quantum theoretical model, the measured standard deviation (SD) of the time offset $\Delta t_0$ can be given by the detected coincidence width $\sigma$ of the entangled photons and the number of photon pairs $N$ within a certain measurement time [13]:

$$\langle \Delta t_0 \rangle_N = \sigma/\sqrt{N} \tag{1}$$

where the detected coincidence width $\sigma$ involves the dispersion broadening of the fiber link and the timing jitter of the single photon detectors (about 51 ps for our SNSPDs). In the experiment, the coincidence widths were measured to be about 132 ps for both $t_2 - t_1$ and $t_3 - t_4$, which is close to the theoretical value of 121 ps. Due to the low sampling rate of the utilized event timers (~6 kHz for each port), the acquired number of photon pairs $N$ in 30 s was only around 1440 cps. By substituting these two parameters into Eqn. (1), theoretical estimation of 1.7 ps was derived and consistent with the experimental result. From the above analysis, the improvement of $\Delta t_0$ for two orders of magnitude can be expected by narrowing the coincidence width with NDC optimization [24] and utilizing new ETs with MHz-level sampling rate.

With the increasing of the average time, the TDEV curve reaches the inflexion at about 1000 s instead of 2000 s, indicating the effect of the frequency transfer technology on suppressing the influence of the independent circumstance temperature fluctuations. The long-term time stability approaches 5.2 ps at 7680 s, which shows an improvement by 4 times. The drop trend with the measurement time promises better time stability at longer averaging time beyond 10000 s.

**Conclusions**

In summary, we have demonstrated a field two-way quantum synchronization of distant clocks. Over a 7 km-long deployed telecommunication fiber link, the H maser at NTSC and the Rb clock at LSO are compared and synchronized via the two-way quantum clock synchronization method. Analogous to the previous report [14], the short-term synchronization stability reached the fundamental limit determined by the relative frequency stability of the Rb clock to the H-maser, which was measured as 32 ps at 30 s. The long-term synchronization stability was achieved to be 19.3 ps at 7680 s. Further applying the fiber-based microwave frequency-transfer to the experiment, a significantly improved time stability was achieved with a short-term TDEV of 1.9 ps at 30 s and a long-term TDEV of 5.2 ps at 7680 s. The nice agreement between the short-term TDEV result and theoretical analysis proves that, both the low sampling rates of the utilized event timers (ET, ~6 kHz) for coincidence measurement and imperfect NDC limit the achievable stability. Utilizing new ETs with MHz sampling rate and optimizing the NDC effect, a magnificent advancement of the stability to the level set by the transferred frequency stability can be achieved. This experiment proves the high potential of the two-way quantum clock synchronization in field applications to promote the synchronization performance. For example, our method can be easily applied to the time-bin encoded quantum key distribution (QKD),to improve the synchronization required for much enhanced key rate.

**Acknowledgement**

This work was supported by the National Natural Science Foundation of China

(Grant Nos. 12033007, 61801458, 61875205, 91836301), the Frontier Science Key Research Project of Chinese Academy of Sciences (Grant No. QYZDB-SW-SLH007), the Strategic Priority Research Program of CAS (Grant No. XDC07020200), the Youth Innovation Promotion Association, CAS (Grant No. 2021408), the Western Young Scholar Project of CAS (Grant Nos. XAB2019B17 and XAB2019B15), the Chinese Academy of Sciences Key Project (Grant No. ZDRW-KT-2019-1-0103), and the West Light Foundation of the Chinese Academy of Sciences (Grant No. 29202082).

**Reference:**

[1] Petit G. and Jiang Z., Precise point positioning for TAI computation, Int. J. Nav. Obs. 2008, 562878 (2008).

[2] Parker T. E. and Zhang V., Sources of instabilities in two-way satellite time transfer, in Proceedings IEEE International Frequency Control Symposium and Exposition, (Institute of Electrical and Electronics Engineers, New York, 2005), pp. 745-751.

[3] Guillemot, P. H., Samain, E., Vrancken, P., Exertier, P., and Leon, S. Time transfer by laser link-T2L2: an opportunity to calibrate rf links. Proc. of the PTTI 2008 95-106 (2008)

[4] Laas-Bourez, M., Courde C., Samain E., Exertier P., Guillemot P., Torre J. M., Martin N., and Foussard C. Accuracy validation of T2L2 time transfer in co-location. IEEE T. Ultrason. Ferr. 62 255-265 (2015)

[5] Guillemot P., Gasc K., Petitbon I., Samain E., Vrancken P., Weick J., Albanese D., Para F., and Torre J. M., Time transfer by laser link: The T2L2 experiment on Jason 2. In International Frequency Control Symposium and Exposition, 2006 IEEE, pages 771–778, (2006).

[6] Samain E., Exertier P., Guillemot P., Pierron F., Albanese D., Paris J., Torre J., and Leon S., Time transfer by laser link - T2L2: Current status of the validation program. In EFTF-2010 24th European Frequency and Time Forum, pages 1–8, (2010).

[7] Lopez O., Kanj A., Pottie P.-E., Rovera D., Achkar J., Chardonnet C., Amy-Klein A., and Santarelli G., Simultaneous remote transfer of accurate timing and optical frequency over a public fiber network, Appl. Phys. B 110, 3-6 (2012).

[8] M. Rost, D. Piester, W. Yang, T. Feldmann, T. Wübbena, and A. Bauch, Time transfer through optical fibres over a distance of 73 km with an uncertainty below 100 ps, Metrologia 49, 772 (2012).

[9] Bothwell, T., Kedar D., Oelker E., Robinson J. M., Bromley S. L., Tew W. L., Ye J. and Kennedy C. J., JILA SrI optical lattice clock with uncertainty of $2.0 \times 10^{-18}$. Metrologia, 56, 065004 (2019).

[10] Yudin V. I., Taichenachev A. V., Basalaev M. Y., Prudnikov O. N., Fürst H. A., ehlstäubler T. E. and Bagayev S. N., Combined atomic clock with blackbody-radiation-shift-induced instability below $10^{-19}$ under natural environment conditions, New J. Phys. 23 023032 (2021)

[11] Hou F., Dong R., Quan R., Xiang X., Liu T., Zhang S., First Demonstration of nonlocal two-way quantum clock synchronization on fiber link, Conference on Lasers and Electro-Optics/Pacific Rim (pp. Th4J-3). Optical Society of America, (2018)

[12] Hou F., Quan R., Dong R., Xiang X., Li B., Liu T., Yang X., Li H., You L., Wang Z., and Zhang S., Fiber-optic two-way quantum time transfer with frequency-entangled pulses, Phys. Rev. A 100, 023849 (2019)

[13] Lee J., Shen L., Cerè A., Troupe J., Lamas-Linares A., and Kurtsiefer C., Symmetrical clock synchronization with time-correlated photon pairs, Appl. Phys. Lett. 114, 101102 (2019)

[14] Samain E., An Ultra Stable Event Timer, Proceedings of the 13th International Workshop on laser ranging instrumentation, 2002.

[15]. Islam, N. T., Lim, C. C. W., Cahall, C., Kim, J., and Gauthier, D. J., Provably secure and high-rate quantum key distribution with time-bin qudits, Science advances 3 (11), e1701491 (2017).

[16]. Xu Liu, Xin Yao, Heqing Wang, Hao Li, Zhen Wang, Lixing You, Yidong Huang, and Wei Zhang, Energy-time entanglement-based dispersive optics quantum key distribution over optical fibers of 20 km, Appl. Phys. Lett. 114, 141104 (2019)

[17] Zhang Y., Hou F., Liu T., Zhang X., Zhang S., Dong R., Generation and quantum characterization of miniaturized frequency entangled source in telecommunication band based on type-II periodically poled lithium niobate waveguide, Acta Phys. Sin. 67, 14 (2018)

[18] You L., Yang X., He Y., Zhang W., Liu D., Zhang W., Zhang L., Zhang L., Liu X., Chen S., Wang Z., and Xie X., "Jitter analysis of a superconducting nanowire single photon detector," AIP Adv. 3(7), 072135 (2013).

[19] Zhou H., He Y., You L., Chen S., Zhang W., Wu J., Wang Z., and Xie X., "Few photon imaging at 1550 nm using a low-timing-jitter superconducting nanowire single-photon detector," Opt. Express 23(11), 14603–14611 (2015).

[20] Quan R., Dong R., Xiang X., Li B., Liu T., and Zhang S., High-precision nonlocal temporal

correlation identification of entangled photon pairs for quantum clock synchronization, Rev. Sci. Instrum. 91, 123109 (2020)

[21] Galleani L., A tutorial on the two-state model of the atomic clock noise, Metrologia, 45, S175–S182, (2008)

[22] Li B., Hou F., Quan R., Dong R., You L., Li H., Xiang X., Liu T., and Zhang S., Nonlocality test of energy-time entanglement via nonlocal dispersion cancellation with nonlocal detection, Phys. Rev. A, 100, 5, (2019)

[23] Xue W., Zhao W., Quan H., Zhao C., Xing Y., Jiang H. and Zhang S., Microwave frequency transfer over a 112-km urban fiber link based on electronic phase compensation, Chin. Phys. B, 29, 064209, (2020)

[24] Xiang, X., Dong, R., Li, B., Hou, F., Quan, R., Liu, T., and Zhang, S. Quantification of nonlocal dispersion cancellation for finite frequency entanglement. Opt. Express, 28(12), 17697-17707 (2020).